\documentclass[12pt]{iopart}

\usepackage{booktabs}
\usepackage{enumitem} 
\usepackage[a4paper, left=1in, right=1in, top=1in]{geometry}
\usepackage{graphicx}
\usepackage{iopams}
\usepackage{multirow}
\usepackage{subcaption}
\usepackage{xcolor}

\newcommand{\blockcomment}[1]{}

\begin{document}
\title[Impact of Forecast Stability on Navigational Contrail Avoidance]{Impact of Forecast Stability on Navigational Contrail Avoidance}

\author{T R Dean$^{1,\dagger,*}$, T H Abbott$^{1,\dagger}$, Z Engberg$^1$, N Masson$^1$, R Teoh$^2$, J P Itcovitz$^2$, M E J Stettler$^2$, M L Shapiro$^1$}

\address{$^1$ Breakthrough Energy, Kirkland, WA, United States of America}
\address{$^2$ Department of Civil and Environmental Engineering, Imperial College London, London, United Kingdom}
\address{$^\dagger$ Denotes equal contributions by authors.}
\address{$^*$ Author to whom any correspondence should be addressed.}
\ead{thomas.dean@breakthroughenergy.org}
\begin{abstract}
Mitigating contrail-induced warming by re-routing flights around contrail-forming regions requires accurate and stable forecasts of the state of the upper troposphere and lower stratosphere. Forecast stability (i.e., consistency between forecast cycles with different lead times) is particularly important for ``pre-tactical'' contrail avoidance strategies that adjust routes based on forecasts with lead times as long as 24-48 hours. However, no study to date has systematically quantified the degree to which forecast stability limits the effectiveness of pre-tactical avoidance. This study addresses this gap by comparing contrail forecasts generated using ECMWF HRES weather forecasts with lead times up to 48 hours to contrail hindcasts generated based on ECMWF ERA5 reanalysis. An analysis of forecast errors show low pointwise consistency between persistent-contrail-forming regions in forecasts and reanalysis, with pointwise error rates similar to those found in previous comparisons of contrail-forming regions in reanalysis and reality. However, we also show that spatial errors in the locations of contrail-forming regions are relatively small, both when forecasts are compared to reanalysis and when reanalysis is compared to in-situ measurements. Finally, we show that designing a trajectory optimizer to take advantage of relatively small spatial errors allows flight trajectory optimizations based on contrail forecasts to reduce contrail climate forcing evaluated based on reanalysis by 80-90\% at the 8-24 hour lead times most relevant to flight planning, with fuel penalties under 0.4\%. Our results show that forecasts with lead times relevant to flight planning are stable enough to be used for pre-tactical contrail avoidance. 

\end{abstract}
\vspace{2pc}

\section{Introduction}
Persistent contrails and contrail-induced cirrus clouds are responsible for about half of the current anthropogenic climate forcing from aviation \cite{lee_2021_contribution}.  Contrails are ice clouds that can form when the jet engine exhaust plume, which consists of combustion byproducts including particulate matter and water vapor, mixes with cold ambient air and becomes supersaturated with respect to water.  This occurs when the Schmidt-Appleman criterion  (SAC) is satisfied (i.e., when the ambient temperature falls below the SAC threshold temperature), causing water vapor to condense onto the surface of particles found in the exhaust plume and subsequently freeze to form ice crystals \cite{schumann_conditions_1996, karcher_role_2009,yu_2024_revisiting}.  If the ambient atmosphere is supersaturated with respect to ice (i.e., relative humidity with respect to ice, RHi, is above 100\%), then the ice particles will persist and the contrail will spread, over time becoming indistinguishable from natural cirrus clouds \cite{haywood_case_2009, teoh_high-resolution_2024, vazquez-navarro_contrail_2015}.  Estimates of current effective radiative forcing from persistent contrails and contrail cirrus are comparable to, and in some cases higher than, estimates of radiative forcing from historical aviation CO2 emissions  \cite{lee_2021_contribution, meerkotter_radiative_1999, karcher_formation_2018, schumann_life_2017}.

There are several proposed solutions to mitigate the climate impacts of contrails.  Some studies have suggested that contrails formed by flights using sustainable aviation fuel (SAF) may be optically thinner, shorter-lived, and less strongly warming than contrails formed by flights using conventional jet fuels \cite{caiazzo_impact_2017, teoh_targeted_2022, voigt_cleaner_2021}. However, recent studies have identified several challenges in using SAF to reduce contrail climate forcing. The costs of SAF-based contrail abatement ($\mathrm{EUR}\,14$-61/tCO2e) are likely significantly higher than the costs of navigational avoidance ($< \mathrm{EUR}\,1$-3/tCO2e) after accounting SAF supply chain costs \cite{woeldgen2025sustainable}. Additionally, SAF availability remains low ($\sim0.2\%$ of the global annual aviation fuel consumption as of 2023) with slow growth rates \cite{watson2024sustainable, frias_feasibility_2024}. Thus, while SAF may play an important role in decarbonizing aviation, its potential for near-term contrail mitigation is limited.  

Another proposed mitigation strategy is the use of lean-burn combustors as an engine option \cite{teoh_mitigating_2020}. Lean-burn combustors emit up to five orders of magnitude less non-volatile particulate matter (nvPM) than conventional engines \cite{icao_edb} and could, in principle, produce optically-thinner and shorter-lived contrails if water droplets nucleated exclusively on nvPM. However, recent in-flight measurements reported initial apparent contrail ice emissions index on the order of $10^{14}$ to $10^{15}$ kg$^{-1}$, comparable to those from conventional engines, despite the nvPM number emissions index being only around $10^{11}$ kg$^{-1}$. This discrepancy may be explained by the presence of volatile particulate matter (vPM, i.e., a mixture of engine lubrication oil droplets, sulfuric acid, and organic compounds), which likely serve as condensation nuclei when nvPM is scarce \cite{yu_2024_revisiting, voigt2025substantial,  ponsonby2024jet, ponsonby2025updated}.

A third approach to mitigating the climate impact of contrails is through navigational avoidance of regions where warming contrails can form \cite{frias_feasibility_2024,schumann_2011_potential, grewe_feasibility_2017, sonabend2024feasibility, simorgh_2025_climate}.  The simplest form of navigational avoidance---avoidance of ice super-saturated regions (ISSRs) that support persistent contrails---may be feasible, but could require avoiding a prohibitively large volume of airspace, as quantified in \cite{engberg_2025_forecasting}.  However, significant contrail mitigation is still possible without avoiding all ISSRs, as around 80\% of contrails do not persist beyond 5 minutes and around 30\% of persistent contrails have a net cooling effect \cite{teoh_high-resolution_2024}. Accordingly, several studies have proposed using forecasts of contrail lifetime and radiative forcing (generated e.g. based on the contrail cirrus prediction model, or CoCiP) to avoid only the subset of ISSRs where particularly long-lived or strongly-warming contrails form \cite{teoh_beyond_2020, schumann_contrail_2012, engberg_2025_forecasting}.

Navigational avoidance strategies will likely cause some amount of increased fuel consumption and a proportional increase CO$_2$ emissions \cite{frias_feasibility_2024}.  Due to the differing nature of the lifetime and uncertainties associated with each of these forcing agents, quantifying overall climate costs and benefits associated with a contrail mitigation scheme is a complex topic and outside of the scope of this work. However, we note that recent studies have suggested that navigational avoidance of contrails is highly likely to have a positive climate outcome despite uncertainties about the magnitude of contrail climate impacts and the most appropriate choice of impact metric \cite{prather2025trade,smith2025climate, borella2024importance}.

This paper focuses specifically on ``pre-tactical'' avoidance, where contrail-aware flight plans are generated at the time of flight planning.  
This approach differs from ``strategic'' avoidance, which relies on climatological properties of contrails (e.g., by reducing flight distance at night when contrails are almost always net warming), and ``tactical'' avoidance, which relies on requesting mid-flight trajectory changes to avoid contrail-forming regions.  
For commercial airlines with regularly-scheduled flights, optimized flight trajectories are typically generated by flight planning software within 24 hours of the flight's departure. 
Routes are optimized according to a set of criteria that will vary by operator, but will include some combination of minimizing fuel, time, and other operational costs. Optimized trajectories are manually adjusted by flight crews and air traffic controllers to account for en-route hazards and other operational issues such as airspace congestion.
Depending on the demands of the operator, flight trajectories may be re-optimized within a few hours departure using the most up-to-date weather forecasts. 
Flight planning at lead times exceeding 24 hours may occur in special situations, such as planning for cargo capacity on long-haul routes, or due to delays in forecast delivery.

Simulation-based studies have shown that, given accurate forecasts of contrail warming, pre-tactical contrail avoidance may be able to achieve a 70\% reduction in contrail radiative forcing with only a $\sim0.1\%$ increase in fleet-wide fuel consumption \cite{frias_feasibility_2024}. However, it remains unclear whether forecasts of upper-tropospheric and lower-stratospheric (UTLS) water vapor and temperature are sufficiently accurate to avoid warming contrails in practice. Previous studies comparing forecasts and reanalysis to in-situ observations from aircraft and radiosondes have found that forecasts and reanalysis products exhibit poor pointwise agreement with observations, with equitable threat scores below 0.4 \cite{gierens_how_2020, agarwal_reanalysis-driven_2022, hofer_well_2024, thompson_fidelity_2024}. These results have been used to argue that pre-tactical contrail avoidance may have limited climate benefits.

In addition to accurate forecasts, navigational avoidance at the flight planning stage requires stable forecasts that change relatively little over the (typically 8-24 hour) period between flight planning and departure.  A high degree of forecast stability would allow planners flexibility in choosing the forecast cycle used for trajectory optimization. In contrast, low forecast stability would force planners to choose between strongly distinct routes based on different forecast cycles, potentially with little guidance as to which will produce the least contrail warming when flown.  While previous studies clearly illustrate potential risks posed by unstable forecasts \cite{molloy2022design}, no study to date has systematically examined the degree to which forecast stability limits the effectiveness of navigational avoidance.

In this study, we quantify the impact of forecast stability on the effectiveness of pre-tactical contrail avoidance by 
\begin{enumerate}
\item generating pairs of cost- and contrail-optimal trajectories based on contrail forecasts with a range of lead times, and
\item evaluating the difference in contrail energy forcing (i.e., lifetime-integrated contrail radiative forcing) between cost- and contrail-optimal trajectory pairs using reanalysis-based contrail hindcasts.
\end{enumerate}
Weather forecasts are initialized from meteorological state estimates (``analyses") produced by data assimilation systems that combine model outputs with observations. Because forecasts are free-running, they include errors that grow with increasing lead time. In contrast, reanalyses are continuously constrained by assimilated observations, and optimizing trajectories using forecasts and evaluating reductions in contrail warming using reanalysis therefore provides a way to examine the impact of forecast error growth. (It also accounts for the impact of errors in analyses, which are required to be available in near-real-time, relative to reanalysis, which is produced retrospectively and can assimilate a larger set of observations.)
We emphasize, however, that this approach does not provide direct insight into the impact of forecast accuracy (i.e., differences between forecasts and reality) on the effectiveness of navigational avoidance because reanalysis is not a perfect source of meteorological truth.

The remainder of the paper proceeds as follows. In Section \ref{sec:methods}, we provide background on methods used to generate contrail forecasts and analyze forecast errors. In Section \ref{sec:design}, we provide a preliminary analysis of errors in contrail-forming regions in forecasts relative to reanalysis and reanalysis relative to in-situ measurements, and describe how this analysis informed the design of the trajectory optimizer used to generate cost- and contrail-optimal trajectories. In Section \ref{sec:results}, we present the primary result of this study and show that contrail-optimal trajectories generated using forecasts can produce large (70-100\%) reductions in contrail warming evaluated using reanalysis. We provide some additional analysis of the behavior of our optimizer in Section \ref{sec:discussion} and finally offer outlooks and conclusions in Section \ref{sec:conclusion}.

\section{Background}
\label{sec:methods}

\subsection{Meteorology data}

We used the European Center for Medium-Range Weather Forecasts (ECMWF) IFS HRES as our source of weather forecasts and the ECMWF ERA5 HRES reanalysis \cite{hersbach_era5_2020} as our source of reanalysis data\footnote{Note that this manuscript uses ``forecast'' or ``HRES'' to refer to ECMWF IFS HRES forecasts and ``reanalysis`` or ``ERA5'' to refer to ECMWF ERA5 HRES reanalysis.}. We retrieved forecast and reanalysis fields on model levels at 0.25 degree horizontal resolution. Unless otherwise noted, forecast fields were retrieved from the ECMWF Operational Archive and reanalysis fields from the Copernicus Climate Data Store. In all cases, we preprocessed forecast and reanalysis data before further use by interpolating fields onto 30 pressure levels roughly equally spaced between FL200 and FL500 and by applying the quantile-matching humidity scaling described by \cite{platt_effect_2024} to forecast and reanalysis specific humidity.

\subsection{Contrail forecasts}

We generated forecasts of contrail warming using the CoCiP grid model \cite{engberg_2025_forecasting} with HRES forecasts as input.  The CoCiP grid model is an approximation of the original ``trajectory'' version of CoCiP \cite{schumann_contrail_2012}, a Lagrangian Gaussian plume model that estimates contrail properties for complete flight trajectories. Rather than estimating properties for full flight trajectories, CoCiP grid simulates the evolution of infinitesimal contrail segments initialized on a 4D spatiotemporal grid to produce a gridded estimate of contrail energy forcing (EF) produced by flights through different volumes of airspace.  This requires making simplifying assumptions about aircraft performance and contrail orientation (see \cite{engberg_2025_forecasting} for details), but allows CoCiP grid to efficiently compute global forecasts of contrail EF in a format that can be easily integrated into commercial flight planning software. The metric of per-waypoint EF is defined using the same conventions as the original formulation of CoCiP \cite{schumann_contrail_2012}, which is obtained by integrated the radiative flux across the area and lifetime of contrail segment formed between a flight waypoint and its subsequent waypoint.

\begin{table}[t!]
    \centering
    \renewcommand{\arraystretch}{1.2}
    \begin{tabular}{c|c|c}
        \toprule
         Emissions Category & Aircraft Types & Representative Aircraft Class \\
         \midrule
         High-nvPM & A20N, A21N & A320 + IAE V2527-A5 \\
         \multirow{2}{*}{Low-nvPM} & \multirow{2}{*}{\shortstack{A19N, B38M, B748 \\ B788, B789, B78X}} & \multirow{2}{*}{B789 + GEnx-1B76} \\
         &  &  \\
         Nominal-nvPM & All others & B738 + CFM56-7B26 \\
         \bottomrule
    \end{tabular}
    \caption{Mapping from ICAO aircraft types to the aircraft classes used as representative in CoCiP grid forecasts for their emissions types.}
    \label{tab:classes}
\end{table}
EF estimates from CoCiP grid are sensitive to the choice of aircraft type and engine model.  However, it is possible to reduce the computational costs associated with generating forecasts for a large number of aircraft and engine types by grouping similar aircraft into a small number of classes, where each class includes aircraft-engine combinations that produce similar contrail climate impacts \cite{engberg_2025_forecasting}.  In this study, we used three aircraft classes to represent aircraft/engine combinations with different levels of nvPM emissions (see also Table \ref{tab:classes}):
\begin{itemize}
    \item{Low-nvPM Group: B789 equipped with GEnx-1B76 engine,}
    \item{Nominal-nvPM Group: B738 equipped with CFM56-7B26 engine,}
    \item{High-nvPM Group: A320 equipped with IAE V2527-A5 engine.}
\end{itemize}
Gridded forecasts for each aircraft class were generated following \cite{engberg_2025_forecasting}, using performance and emissions calculations for the representative aircraft-engine combination.

Throughout this work, we use CoCiP grid to estimate contrail impacts when performing trajectory optimization. When possible, contrail impacts of an individual flight trajectory are computed using the higher-fidelity trajectory version of CoCiP, accounting for the correct aircraft type, and engine type. The trajectory version of CoCiP estimates fuel flow rates across the aircraft trajectory using an assumed load fact, providing more accurate estimates of nvPM and initial ice concentration numbers than the grid variant of CoCiP which relies on nominal engine efficiency values to estimate fuel consumption.  

\subsection{Forecast error analysis}

We quantified the frequency of pointwise errors in contrail forecasts using the equitable threat score (ETS) metric \cite{gierens_how_2020,hofer_well_2024}.
Unless otherwise noted, we use ``pointwise'' to denote metrics computed per flight waypoint.
For a set of predictions of a binary variable with $a$ true positives, $b$ false positives, $c$ false negatives, and $d$ true negatives, the ETS is
\begin{equation}
    ETS = \frac{a - r}{a + b + c - r}\,\,,
\end{equation}
where $r = (a + b)(a + c)/(a + b + c + d)$ is the expected number of true positives produced by random matches.  The ETS for a random model is 0 and the ETS for a perfect model is 1. With a low base rate (far fewer positives than negatives), the ETS for a model with precision and recall of 0.5 is $1/3$. To ensure that ETS scores reflected the frequency of errors at locations relevant to flight planning, we computed ETS scores from predictions interpolated to flight waypoints.

Pointwise measures of contrail forecast accuracy are vulnerable to the so-called ``double-penalty'' issue, widely discussed in the literature on evaluating high-resolution forecasts (e.g.,~\cite{ebert_2008_fuzzy}). Because contrail forecasts predict localized features, relatively small displacement errors in the predicted locations of contrail-forming regions can produce very frequent pointwise errors. To address this limitation, we also quantified the proximity of forecasted contrail-forming regions to contrail-forming regions found in reanalysis based on a ``proximity distribution''. To compute proximity distributions, we took sets of flight trajectories, calculated the flight time from each flight waypoint to the nearest forecasted contrail-forming region along the same flight trajectory (assigning a flight time of infinity for waypoints on flight trajectories for which no contrails were forecasted), and computed probability distributions from that set of flight times. The proximity distribution for flight waypoints in true contrail-forming regions provides a measure of their proximity to contrail-forming regions in forecasts. We note that because flights follow approximately horizontal trajectories, this metric primarily captures horizontal rather than vertical proximity.

\subsection{Statistical Methods} \label{sec:methods:stats}
Unless otherwise stated, statistics presented as a function of forecast lead time were computed by binning waypoints based on a single nominal lead time assigned to all waypoints in a flight. Nominal lead times were computed as the difference between the forecast initialization time and the time of the flight's first waypoint. Error bars for binned statistics correspond to 95\% confidence intervals obtained by applying a bias-corrected accelerated (BCa) bootstrap analysis \cite{efron1987better}, which we now briefly describe.  For each statistic, bootstrap samples are obtained by resampling (with replacement and before binning) from the set of all flights considered in this study (i.e., they are resampled per flight and not per day or per waypoint). Each statistic is resampled a total of 10,000 times.
The empirical collection of resampled statistics is known as a bootstrap distribution. The confidence intervals are chosen from the tails of this bootstrap distribution. 
In the most basic form of bootstrap, the confidence interval based on quantiles of the bootstrap distribution.
However, if the bootstrap distribution is biased or skewed, then this method will not produce an accurate estimate of the confidence intervals. In this case, they may be chosen according the BCa method originally described by \cite{efron1987better}, which accounts for both bias and skewness in the bootstrap distribution.

\section{Optimizer design} \label{sec:design}

The optimizer used in this work is similar to the optimizer described in Appendix A6 of \cite{engberg_2025_forecasting}. It requires an existing aircraft trajectory as input, but retains only the takeoff time and the latitude and longitude of each waypoint. The space along the horizontal trajectory is divided into a two-dimensional grid, with the horizontal dimension representing distance along the flight path and the vertical dimension representing altitude.  The horizontal dimension is divided into equally spaced points that are $\sim 1\,\mathrm{minute}$ apart at a nominal cruise speed for the aircraft class, obtained from aircraft performance models \cite{poll_estimation_2021}. The vertical dimension is divided into standard flight levels (1000 feet increments). The time at which the flight crosses each horizontal point is assumed to be independent of the choice of vertical trajectory when meteorological fields and EF from gridded CoCiP are interpolated onto this two-dimensional grid.  Because only the vertical trajectory is re-optimized, this optimizer simulates navigational avoidance based solely on vertical deviations (i.e., no lateral deviations).

\begin{figure}[t!]
    \centering
    \includegraphics[width=\linewidth]{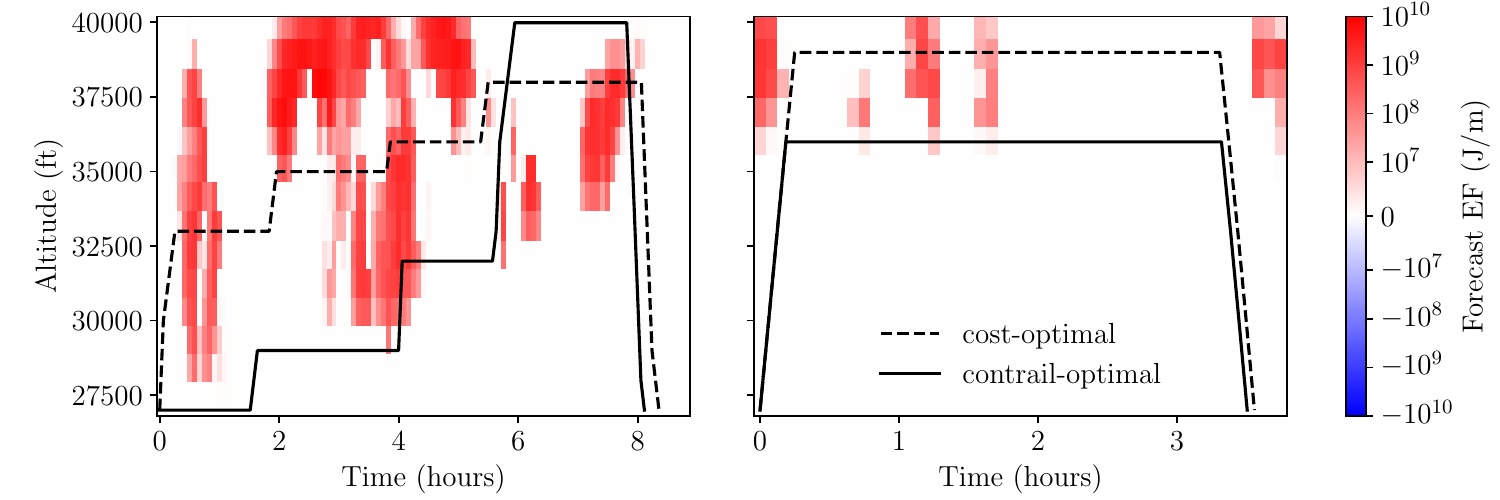}
    \caption{Two example optimized flight trajectories, showing for each a cost-optimal trajectory (dashed line) and a contrail-optimal trajectory (solid line). The trajectory on the left-hand side reduces contrail EF by $1.2 \times 10^{15} \,\mathrm{J}$ ($340,000 \,\mathrm{kg}$ CO$_2$e\textsubscript{AGWP100}), with an additional fuel burn of $2,233 \,\mathrm{kg}$ ($+4.7\%$). The trajectory on the right-hand side saves $6.3 \times 10^{13} \,\mathrm{J}$ ($18,000 \,\mathrm{kg}$ CO$_2$e\textsubscript{AGWP100}) of contrail EF, with a fuel penalty of $200 \,\mathrm{kg}$ ($+2.4\%$).}
    \label{fig:trajectories}
\end{figure}

The optimizer finds a vertical profile that minimizes the objective function
\begin{eqnarray}
\mathrm{Total\,Cost} &= \mathrm{Fuel\,burn\,} [kg] \label{eqn:Optimizer Cost} \\
&+ \mathrm{Cost\,Index\,} [kg / min] * \mathrm{Flight\,time\,} [min] \nonumber \\
&+ \mathrm{\,Contrail\,Cost\,Index\,}[kg / J] * \mathrm{Contrail\,EF\,}[J] \nonumber. 
\end{eqnarray}
This is accomplished through a breadth-first Dykstra-like search across the 2D grid described above.  Precisely, the optimizer explores all feasible vertical trajectories (search paths) across the 2D grid in order of increasing total cost, and returns the lowest-cost trajectory. A more complete description of the optimization procedure is described in Appendix A6 of \cite{engberg_2025_forecasting}.
Fuel burn was estimated using the Poll-Schumann aircraft performance model \cite{poll_estimation_2021, poll_estimation_2021-1}, as implemented in \cite{shapiro_pycontrails_2023}. The Poll-Schumann model also provides estimates of thrust limitations and flight envelope restrictions (i.e., minimum and maximum permitted Mach numbers), and we ensured that all optimized trajectories satisfied these constraints. We estimated the initial mass of each flight based on an assumed load factor and reserve fuel requirements following \cite{teoh_high-resolution_2024}, and assumed a default engine type based on the mapping from aircraft type provided by \cite{teoh_high-resolution_2024}. We set the cost index to a fixed value of 70 kg/min, effectively assuming that operating an aircraft for one minute costs the equivalent of burning 70 kg of fuel. In practice, this value varies highly between operators and operational conditions. The contrail cost index was set to 0 for cost-optimal trajectories. For contrail-optimal trajectories, the default contrail cost index was set to $6.7\times10^{-11}\,\mathrm{kg/J}$, which converts contrail EF to a CO$_2$-equivalent fuel burn based on an absolute global warming potential over 100 years (AGWP100) \cite{teoh_climate_2020}\footnote{Note that there is not a straightforward relationship between the contrail cost index and the actual fuel penalty paid to fly contrail-optimal rather than cost-optimal trajectories. Using a cost index based on AGWP100 does not imply that the fuel penalty will be equal and opposite to the reduction in contrail warming on AGWP100 terms. The cost index can be tuned to produce different tradeoffs between contrail impacts and costs; a further exploration of this is presented in \ref{sec:fuel}.}. Contrail EF at each waypoint was computed based on a 4D interpolation of gridded CoCiP forecasts for the appropriate aircraft-engine class. Other costs, such as overflight charges or engine cycle costs, were not considered. Example cost- and contrail-optimal trajectories are shown in Figure \ref{fig:trajectories}. It is worth noting that the contrail-optimal trajectories depicted in Figure \ref{fig:trajectories} do not fully avoid contrail-forming regions, but instead find trajectories that balance contrail formation with excess fuel and time costs.

Finally, a preliminary analysis of forecast errors (presented in the following section) motivated the inclusion of an additional constraint on the minimum level flight time required between altitude changes. This constraint is referred to as the ``minimum segment length'' throughout the remainder of the manuscript.

\subsection{Forecast errors and the minimum segment length constraint} \label{sec:design:forecast_errors}

\begin{figure}
    \centering
    \includegraphics[width=0.485\linewidth]{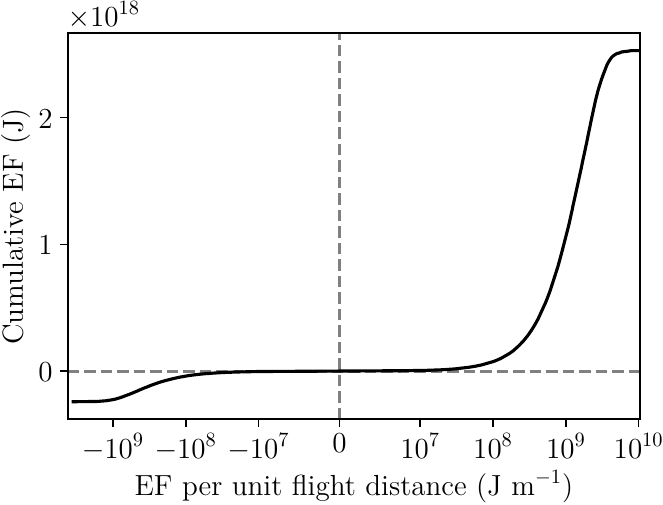}
    \caption{Cumulative contrail EF produced by simulated warming contrails with EF per unit flight distance below a threshold (right side of vertical dashed line) and by simulated cooling contrails with EF per unit flight distance above a threshold (left side of vertical dashed line). EF values shown in this figure were obtained from gridded CoCiP calculations using ERA5 met data interpolated to waypoints along cost-optimal trajectories computed using ERA5 meteorology.}
    \label{fig:ef-cdf}
\end{figure}

The inclusion of a minimum segment length constraint was motivated by an analysis of errors in the locations of contrail-forming regions in HRES forecasts relative to ERA5 reanalysis and in ERA5 reanalysis relative to in-situ measurements. For this comparison of HRES forecasts with ERA5 reanalysis, we used our optimizer to generate three cost-optimal trajectories for each selected flight using HRES forecasts initialized at 00Z the day of the flight, 12Z the day before the flight, and 00Z the day before the flight. A nominal minimum segment length of 90 minutes was used for this re-optimization. We then interpolated fields derived from ERA5 and fields derived from the same HRES forecast used for re-optimization onto waypoints from each re-optimized flight. We considered two variants of a binary contrail forecast for the HRES-ERA5 comparison: one based on ISSR locations and one based on the locations of high-contrail-EF regions using contrail EF computed with gridded CoCiP. We defined high-EF regions as regions with contrail EF per unit flight distance greater than $10^7 \,\mathrm{J/m}$, which captures essentially all contrails that contribute an appreciable amount of warming (Figure \ref{fig:ef-cdf}).

For the analysis of errors in reanalysis relative to in-situ measurements, we compared ERA5 ISSR locations with measurements from the In-service Aircraft for a Global Observing System (IAGOS) dataset \cite{petzold2015global}. We included all IAGOS waypoints from 2019 flights and interpolated ERA5 fields onto those waypoints. ERA5 data processing followed methods described above, except that we retrieved the required year of ERA5 data from the Analysis-Ready Cloud-Optimized ERA5 public dataset \cite{carver2023arco}.

\begin{figure}
    \centering
    \begin{subfigure}{0.485\textwidth}
        \includegraphics[width=\linewidth]{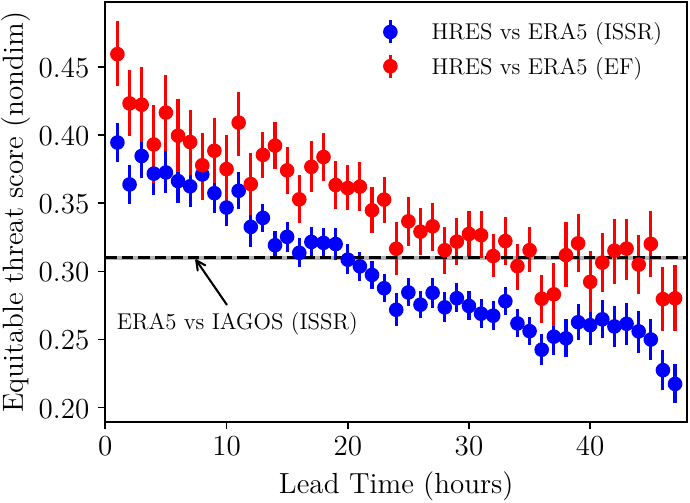}
    \end{subfigure}
    \caption{Equitable threat scores for HRES vs. ERA5 ISSRs (blue), HRES vs. ERA5 high-EF regions (red), and ERA5 vs. IAGOS ISSRs (black). ETS values for HRES vs. ERA5 are shown as a function of forecast lead time, with error bars representing 95\% confidence intervals. The 95\% confidence interval for the ERA5-IAGOS ETS is shown as shaded gray around the dashed black line. High-EF regions are defined as regions with contrail warming per unit flight distance greater than $10^7$ J/m and are a subset of ISSRs.}
    \vspace{-4mm}
    \label{fig:forecast_ets}
\end{figure}

ETS values for ISSR locations in HRES forecasts compared to ERA5 reanalysis are around 0.4 at short lead times and decrease with increasing lead time by about 0.1 per day (Figure \ref{fig:forecast_ets}, blue points). These ETS values are closer to the score for a random model (i.e., ETS = 0) than a perfect model (i.e., ETS = 1) and indicate frequent pointwise disagreement between HRES forecasts of ISSR locations and ISSR locations in reanalysis. At 12-24 hour lead times, ETS values for ISSRs in forecasts relative to reanalysis are similar to ETS values for ISSRs in ERA5 reanalysis relative to IAGOS measurements (Figure \ref{fig:forecast_ets}, black line). ETS values for high-EF regions in forecasts relative to reanalysis are only slightly higher, decreasing from around 0.45 at short lead times to around 0.25 at 48 hour lead times (Figure \ref{fig:forecast_ets}, red points).

\begin{figure}
    \centering
    \begin{subfigure}{0.485\textwidth}
        \includegraphics[width=\linewidth]{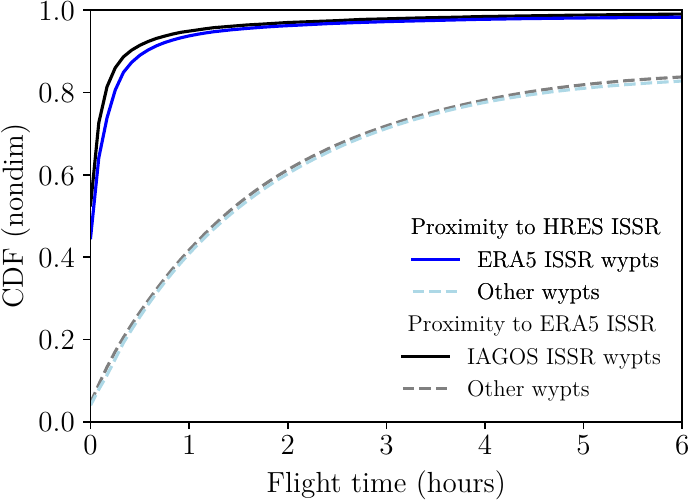}
        \caption{}
        \label{fig:prox_dist_issr}
    \end{subfigure}
    \hfill
    \begin{subfigure}{0.485\linewidth}
        \includegraphics[width=\linewidth]{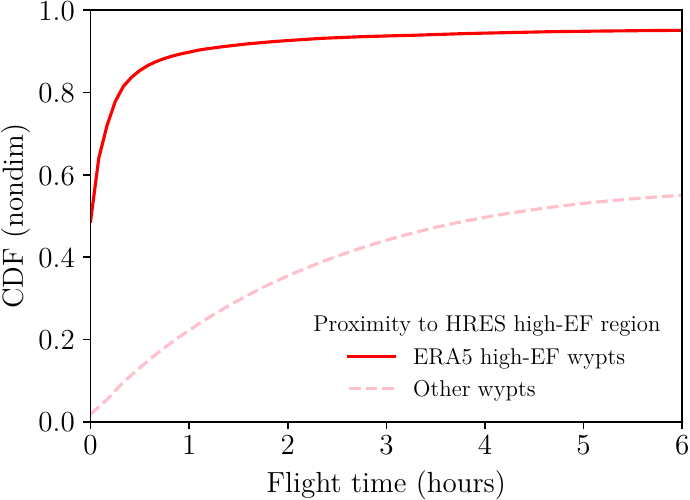}
        \caption{}
        \label{fig:prox_dist_ef}
    \end{subfigure}
    \begin{subfigure}{0.485\linewidth}
        \includegraphics[width=\linewidth]{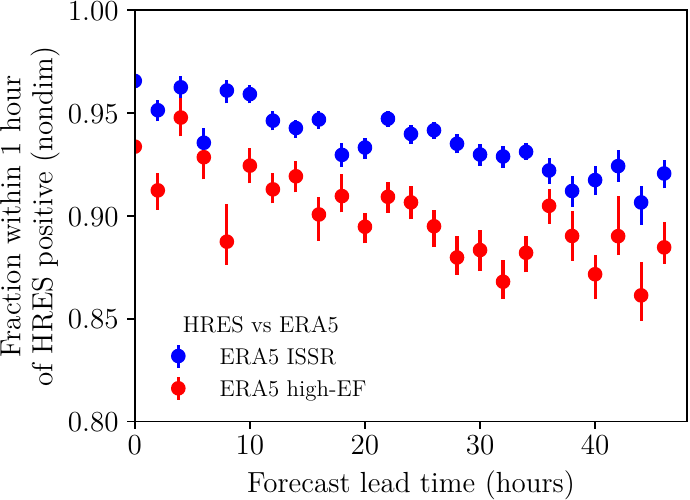}
        \caption{}
        \label{fig:prox_lead}
    \end{subfigure}
    \caption{(a) Cumulative distribution functions (CDFs) measuring the distance between HRES vs. ERA5 ISSRs (blue) and ERA5 vs IAGOS ISSRs (black), henceforth referred to as proximity distributions. (b) Proximity distributions for HRES vs ERA5 high-EF regions. (c) Fraction of waypoints in ERA5 ISSRs that are within an hour's flight time of an HRES ISSR (blue) and of waypoints in ERA5 high-EF regions that are within an hour's flight time of an HRES high-EF region (red), binned by forecast lead time. High EF regions are defined as regions with EF per unit flight distance greater than $10^7$ J m$^{-1}$. Flights with a duration less than four hours are excluded to avoid large differences in the average duration of optimized trajectories and IAGOS flights. In panel (c), ``HRES positives'' refer to ISSRs and high-EF regions for blue and red points, respectively.}
    \vspace{-4mm}
    \label{fig:forecast_proximity}
\end{figure}

 Low ETS scores suggest a large number of pointwise errors between the two products. However, if these pointwise errors are the result of relatively small displacement errors in positions of localized contrail-forming regions, then--just as failing to predict the precise location of a thunderstorm is not necessarily a barrier to deciding to issue a county-level tornado watch--failing to predict the precise locations of contrail-forming regions may not be a barrier to identifying volumes of airspace that flights should avoid. 
 Indeed, ``proximity distributions" of distances between between forecast, reanalysis, and observed contrail-forming regions (Figure \ref{fig:prox_dist_issr}-\ref{fig:prox_dist_ef}) suggest that displacement errors are usually small.
 Waypoints in ERA5 ISSRs and high-EF regions are statistically likely to be close (within 1 hour's flight time) to ISSRs and high-EF regions in HRES forecasts, and waypoints in IAGOS ISSRs are similarly likely to be close to an ERA5 ISSR (Figures \ref{fig:prox_dist_issr}-\ref{fig:prox_dist_ef}, solid lines). This proximity is not merely the result of most waypoints being close to contrail-forming regions: waypoints outside HRES and ERA5 contrail-forming regions are typically much farther from contrail-forming regions in ERA5 and IAGOS measurements, respectively (Figures \ref{fig:prox_dist_issr}-\ref{fig:prox_dist_ef}, dashed lines). The proximity between HRES and ERA5 contrail-forming regions has some dependence on lead time: the fraction of waypoints in ERA5 ISSRs that are within 1 hour's flight time of an HRES ISSR decreases from above 95\% at short lead times to slightly above 90\% at 48 hour lead times, and the equivalent statistic for high-EF regions varies from slightly above 90\% to around 85\% (Figure \ref{fig:prox_lead}).

\begin{figure}
    \centering
    \begin{subfigure}[t]{0.485\linewidth}
        \includegraphics[width=\linewidth]{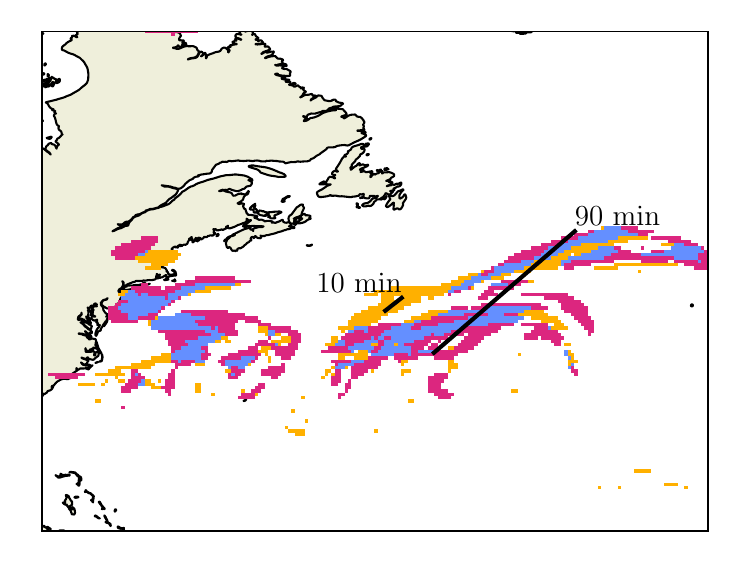}
        \caption{}
        \label{fig:high_ef_map}
    \end{subfigure}
    \hfill
    \begin{subfigure}[t]{0.485\linewidth}
        \includegraphics[width=0.95\linewidth]{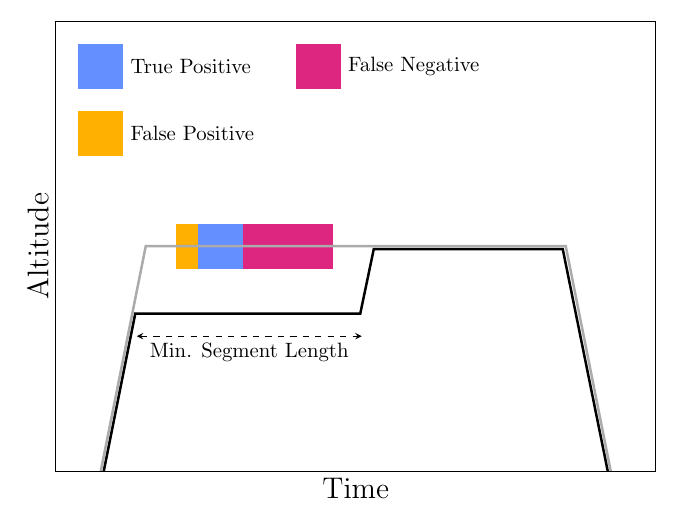}
        \caption{}
        \label{fig:avoidance_schematic}
    \end{subfigure}
    \caption{(a) pointwise agreement between high-EF regions from ERA5 reanalysis and the 36-hour HRES forecast at FL320 at 6Z on 01 March 2019. True positives are blue, false positives are orange, false negatives are red, and true negatives are white. The black lines show the lengths of 90 and 10 minute flight segments assuming a groundspeed of $250 \,\mathrm{m/s}$. (b) Illustration of how trajectory optimization aims to accommodate errors in spatial locations of contrail-forming regions in forecasts and reanalysis by constraining minimum segment lengths between altitude changes. A hypothetical contrail-optimal trajectory (black) has been re-routed from a cost-optimal trajectory (gray) to avoid the regions where forecast EF was high. Due to the constraint on the minimum segment length, the trajectory also avoided a region where forecast EF was low but reanalysis EF was high.}
    \vspace{-4mm}
    \label{fig:min_seg_len}
\end{figure}

Successfully avoiding contrail-forming regions based on imperfect forecasts requires an optimization strategy that takes advantage of forecasts' strengths while limiting the impact of errors. Our analysis suggests that one strength optimizers should take advantage of is spatial proximity between actual and forecasted contrail-forming regions. While a number of different optimization strategies may be capable of taking advantage of small but non-zero errors in forecast locations of contrail-forming regions, the strategy we chose to pursue in this paper was to limit the frequency of altitude changes in optimized trajectories by constraining the minimum flight time required between climbs or descents. Figure \ref{fig:avoidance_schematic} illustrates how this ``minimum segment length'' constraint aims to improve the effectiveness of contrail avoidance. In brief, because actual and forecasted contrail-forming regions are often close to but not exactly coincident with each other, a sufficiently long minimum segment length may increase the probability that contrail avoidance maneuvers planned based on forecasts also avoid contrail-forming regions in reanalysis or reality.

\section{Forecast stability experiments} \label{sec:results}

We used the contrail forecasts described in Section \ref{sec:methods} and the optimizer described in Section \ref{sec:design} to evaluate the impact of lead time-dependent forecast errors on the effectiveness of navigational avoidance. We first selected one random day per month during each month of 2019, and then selected 5000 random commercial flights with takeoff times within the 24 hour period starting at 01Z on each selected day. We chose to select one random day per month---rather than 12 random days throughout 2019---to ensure that we sampled a wide range of meteorological regimes throughout a full seasonal cycle. In \ref{sec:monthly}, we discuss month trends and present evidence to suggest that a larger sample of days is unlikely to change the results presented in this section. We obtained aircraft trajectory information for selected flights from automatic dependent surveillance-broadcast (ADS-B) telemetry purchased from Spire Aviation. 

For each selected flight, we generated three pairs of cost- and contrail-optimal trajectories by re-optimizing the flight's vertical profile based on HRES forecasts initialized at 00Z on the day of the flight, 12Z the day before the flight, and 00Z the day before the flight.  (Note that forecasts influenced cost- as well as contrail-optimal trajectories because temperature and wind forecasts were used for aircraft performance calculations within the optimizer as well as for estimating contrail impacts using CoCiP grid.) Repeating this procedure with a range of minimum segment lengths produced datasets of pairs of optimized trajectories with forecast lead times of 1 to 48 hours. We note that we chose to re-optimize the vertical profiles in the cost-optimal case, rather than retaining the original profiles from the ADS-B data, to eliminate differences between the cost and contrail optimal trajectory caused by, for example, differences in aircraft performance modeling, assumptions about payload weights, or deviations from optimized flight plans.

Given pairs of cost- and contrail-optimal trajectories, we then evaluated the effectiveness of navigational avoidance using contrail EF computed using meteorology data from ERA5 reanalysis.  We emphasize that we based our evaluation on ERA5 reanalysis not because it is a perfect source of meteorological truth---it is not---but rather because it allows us to isolate the impact of forecast error growth on the effectiveness of navigational avoidance. We used ERA5 data for two different versions of evaluation.  For the first version, we computed ERA5 contrail EF using CoCiP grid (following the methods used for forecasts) and interpolated results for the appropriate aircraft class onto optimized trajectories.  For the second version, we computed ERA5 contrail EF using the original ``trajectory'' version of CoCiP run on optimized trajectories. Aircraft performance calculations for trajectory CoCiP used the Poll-Schumann model and made the same assumptions about initial mass and engine type as the optimizer. Results from the first version of evaluation isolate the impact of discrepancies between forecasts and reanalysis, whereas results from the second version also account for the impact of approximations made in CoCiP grid relative to trajectory CoCiP.  Comparing the two versions of evaluation therefore allowed us to determine whether reductions in evaluated contrail EF were limited by disagreement between CoCiP grid and trajectory CoCiP.

\begin{figure}
    \centering
    \begin{subfigure}{0.485\textwidth}
        \includegraphics[width=\linewidth]{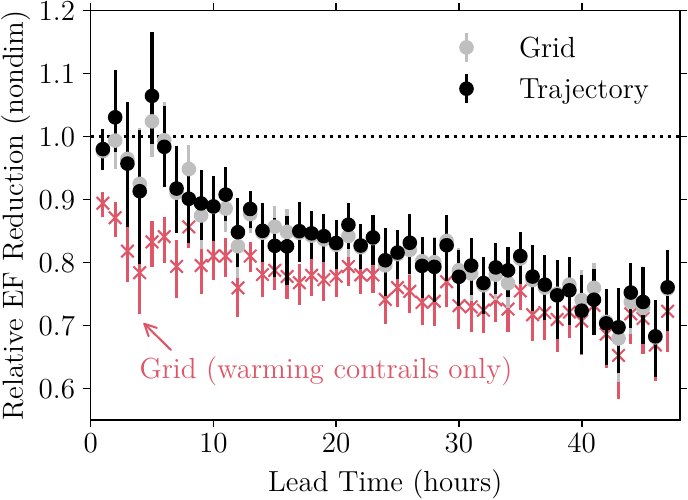}
        \caption{}
        \label{fig:ef_reduction_grid_traj}
    \end{subfigure}
    \hfill
    \begin{subfigure}{0.485\linewidth}
        \includegraphics[width=\linewidth]{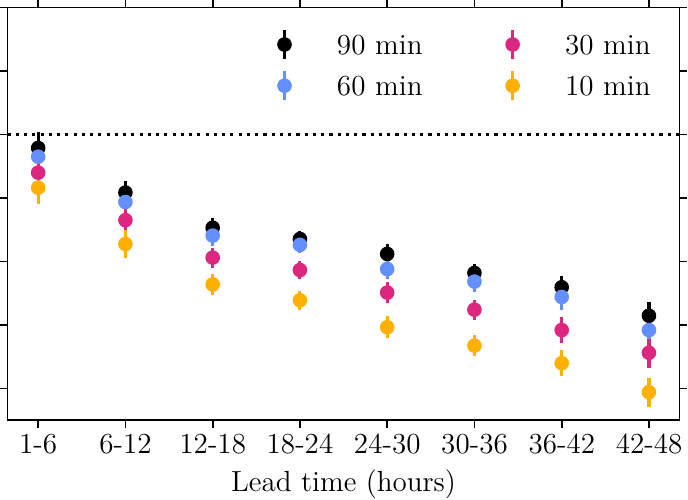}
        \caption{}
        \label{fig:ef_reduction_min_seg_len}
    \end{subfigure}
    \caption{(a) Change in fleet-aggregated contrail EF between contrail- and cost-optimal trajectories as a function of forecast lead time. Trajectories were optimized using HRES forecasts with contrail EF computed using gridded CoCiP, and changes in contrail EF were evaluating using gridded CoCiP (gray) and trajectory CoCiP (black) with ERA5 meteorology. Light red x's show the change in contrail EF evaluated using gridded CoCiP including warming contrails only. (b) Comparison of changes in fleet-aggregated contrail EF obtained using different minimum segment lengths. Only results for evaluation using gridded CoCiP are shown.}
    \vspace{-4mm}
    \label{fig:ef_reduction}
\end{figure}

Figure \ref{fig:ef_reduction_grid_traj} shows the EF reduction, evaluated using ERA5 reanalysis, for contrail-optimal relative to cost-optimal trajectories optimized using HRES forecasts with a minimum segment length of 90 minutes. Forecast errors increase at longer lead times, reducing the fraction of cost-optimal EF avoided by contrail-optimal trajectories. However, the relative EF reduction is high (above 80\%) at the 8-24 hour lead times most relevant for flight planning and remains near 70\% for lead times as long as 48 hours, the longest considered in this study. EF reductions greater than 100\% are reported for some short lead times because cooling contrail segments with negative EF are assigned a contrail cost of zero within the optimizer but included in the calculation of cost- and contrail-EF during evaluation. Recomputing the EF reduction with cooling contrails excluded indicates that contrail-optimal trajectories based on forecasts with short leads times avoid 80-90\% of the EF produced by warming contrails, with all or most of the remaining contrail warming masked by cooling contrails.

For all lead times, EF reductions evaluated using gridded CoCiP match EF reductions computed using trajectory CoCiP to within a few percent. This is consistent with previous work that quantified the pointwise agreement between the two CoCiP variants \cite{engberg_2025_forecasting} and argued that the agreement was likely good enough to allow flight trajectories optimized based on CoCiP grid output to avoid a large fraction of contrail EF calculated by trajectory CoCiP. The close agreement in EF reduction shown here provides quantitative support for that argument and suggests that the gridded version of CoCiP, using three aircraft classes, approximates trajectory CoCiP with sufficient accuracy to be used as a proxy for trajectory CoCiP during flight planning.

Optimizing trajectories using a shorter minimum segment length decreases the fraction of cost-optimal contrail EF avoided by contrail-optimal trajectories (Figure \ref{fig:ef_reduction_min_seg_len}), indicating that a sufficiently long minimum segment length helps to limit the impact of forecast errors. The effect of varying the minimum segment length is particularly pronounced at the longest lead times considered in this study. At 42-48 hour lead times, decreasing the minimum segment length from 90 to 10 minutes leads to a drop in relative EF reduction from above 70\% to below 60\%. The effect is somewhat more modest at the 8-24 hour lead times most relevant for flight planning, but even here reducing the minimum segment length leads to $\sim$10\% decreases in the fraction of contrail EF avoided. We note that a longer minimum segment length also has operational utility in that trajectories with less frequent climbs and descents reduce workload for pilots and air traffic control, and also reduce maintenance costs due to engine wear.

\section{Discussion} \label{sec:discussion}

HRES forecasts with 8-24 hour lead times and ERA5 reanalysis exhibit a poor pointwise agreement on the precise locations where high-warming contrails can form (ETS $<$ 0.4, recall Figure \ref{fig:forecast_ets}). Despite this, we show that trajectory optimization based on HRES forecasts can lead to large (70-100\%) reductions in contrail forcing evaluated using ERA5 reanalysis, particularly when using an optimizer with a minimum segment length constraint tuned to take advantage of spatial proximity of HRES and ERA5 contrail-forming regions (recall Figure \ref{fig:forecast_proximity} and \ref{fig:high_ef_map}). Here, we examine in more detail how our optimzer achieves large EF reductions. While the following discussion focuses on results from our optimizer, it provides a framework that could be used to analyze how the performance of other optimizers is affected by spatiotemporal errors in contrail forecasts.

The framework for this discussion is based on the following premise: in order to successfully avoid a significant fraction of contrail EF, an optimizer
\begin{enumerate}
\item must have a high probability of attempting to avoid cost-optimal flight segments with high EF, regardless whether or not they are forecasted; and
\item must have a high probability of successfully avoiding high-EF regions when an avoidance maneuver is attempted.
\end{enumerate}
In this discussion, we show that our optimizer's minimum segment length constraint aids in accomplishing (i) by increasing the probability that waypoints close to but not within forecast high-EF regions are included in avoidance maneuvers (Section \ref{sec:inclusion}), and we show that our optimizer accomplishes (ii) largely by deviating downward to levels that are too warm for contrail formation (Section \ref{sec:effectiveness}). We conclude by discussing the potential for a similar optimizer to effectively accommodate errors in forecasts relative to reality rather than relative to reanalysis (Section \ref{sec:speculation}).

\subsection{Inclusion in avoidance maneuvers} \label{sec:inclusion}

Our optimizer's minimum segment length constraint is designed to increase the probability that waypoints in high-EF regions close to but not directly coincident with forecast high EF are included in avoidance maneuvers. To show that this constraint has the intended effect, we take all cost-optimal waypoints in ERA5 high-EF regions and bin them by along-trajectory flight distance to the nearest forecast high-EF region, as in the calculation of proximity distributions (Section \ref{sec:design:forecast_errors}). For each bin, we then compute the fraction of the total EF that is associated with waypoints that are included in avoidance maneuvers (i.e., for which the cost- and contrail-optimal altitudes differ).

\begin{figure}[t!]
    \centering
    \begin{subfigure}{0.485\linewidth}
        \includegraphics[width=\linewidth]{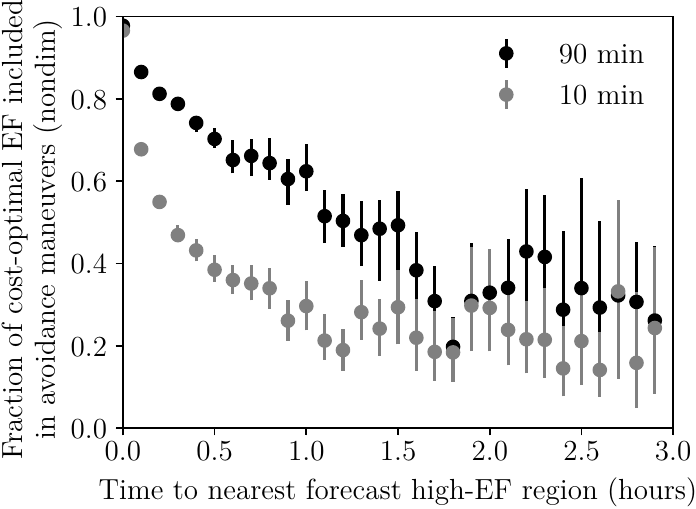}
    \end{subfigure}
    \caption{Fraction of the total cost-optimal EF produced by waypoints in ERA5 high-EF regions that are included in avoidance maneuvers, as a function of flight time to the nearest forecast high-EF region. Black and gray points show fractions for trajectories optimized with minimum segment lengths of 90 and 10 minutes, respectively.}
    \label{fig:avoidance-inclusion}
\end{figure}

The fraction of the total cost-optimal EF produced by waypoints included in avoidance maneuvers is high (close to 1) for waypoints very close to forecast high-EF regions and decreases with increasing flight time from regions of high forecast EF (Figure \ref{fig:avoidance-inclusion}). As intended, using a longer minimum segment length leads to a less rapid decrease. With a minimum segment length of 10 minutes, avoidance maneuvers include only about 30\% of the EF produced by cost-optimal waypoints an hour from a forecast high-EF region, whereas with a minimum segment length of 90 minutes, that figure increases to about 70\%.

\subsection{Effectiveness of avoidance maneuvers} \label{sec:effectiveness}

Re-routing away from locations where EF is high does not, in and of itself, guarantee that avoidance will reduce EF. Avoidance maneuvers must also be ``effective'' in the sense that they re-route trajectories into regions where EF is low. This is important both because it leads to large EF reductions in portions of avoidance maneuvers where cost-optimal EF is high and because it prevents large EF increases in portions of avoidance maneuvers where cost-optimal EF is low. Here, we show that our optimizer's effectiveness is high at short forecast lead times and decreases relatively slowly with increasing lead time. We further show that the high effectiveness is linked to frequent reliance on avoidance maneuvers that target contrail formation rather than persistence.

\begin{figure}[t!]
    \centering
    \begin{subfigure}{0.45\linewidth}
        \includegraphics[width=\linewidth]{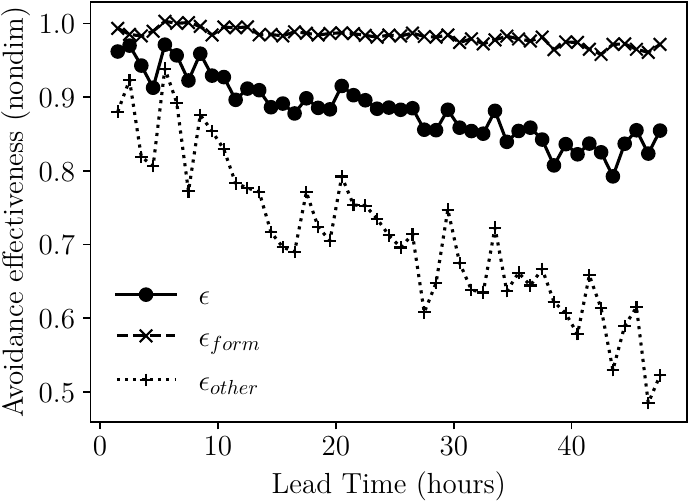}
        \caption{}
    \end{subfigure}
    \hspace{0.3em}
    \begin{subfigure}{0.45\linewidth}
        \includegraphics[width=\linewidth]{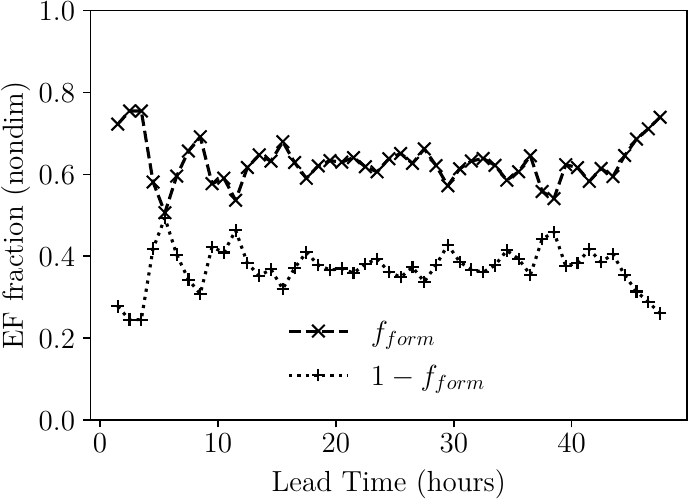}
        \caption{}
    \end{subfigure}
    \caption{(a) Effectiveness of avoidance for all avoidance maneuvers ($\epsilon$) and for avoidance maneuvers that target formation ($\epsilon_{form}$) versus other forecast features ($\epsilon_{other}$). (b) Fraction of EF in avoidance maneuvers that target formation ($f_{form}$), used to relate $\epsilon$ to $\epsilon_{form}$ and $\epsilon_{other}$. See the text for definitions of $\epsilon$, $\epsilon_{form}$, $\epsilon_{other}$, and $f_{form}$.}
    \label{fig:avoidance-success}
\end{figure}

We define the overall ``effectiveness'' of avoidance maneuvers as the ratio of the cost-to-contrail-optimal change in EF to the total cost-optimal EF at waypoints included in avoidance maneuvers. Using $A$ to denote the set of waypoints included in avoidance maneuvers, the effectiveness can be written as
\begin{equation}
    \epsilon = \frac{\sum_A EF_{cost} - EF_{contrail}}{\sum_A EF_{cost}} \,\, ,
\end{equation}
where $EF_{cost}$ is the per-waypoint cost-optimal EF and $EF_{contrail}$ is the per-waypoint contrail-optimal EF. Effectiveness is largest ($\epsilon \sim 1.0$) at short lead times and drops to $\sim 0.8$ at lead times of 48 hours (Figure \ref{fig:avoidance-success}a). We note that $\epsilon$ can in principle exceed 1 as EF is negative at some waypoints (as in Figure \ref{fig:ef_reduction_grid_traj}).

To obtain additional insight into the forecast properties that influence effectiveness, we separate avoidance maneuvers into two categories:
\begin{enumerate}
    \item avoidance maneuvers that attempt to avoid warming contrails by re-routing into regions where contrails are not predicted to form at all, and
    \item avoidance maneuvers that re-route flights into regions where contrails are predicted to form (but may be less strongly warming).
\end{enumerate}
We describe the first category as avoidance maneuvers that ``target formation'', and the second category as avoidance maneuvers that target other features of the forecast (e.g., regions where contrails are predicted to form but not persist, or regions where persistent cooling contrails are predicted). We then use these two categories as the basis for a decomposition of the overall effectiveness into the effectiveness of avoidance maneuvers that target formation and the effectiveness of avoidance maneuvers that target other forecast features. We first define $A_f \subseteq A$ as the subset of waypoints that are re-routed into regions where the SAC for contrail formation is not met in the forecast used for trajectory optimization. SAC satisfaction is determined using a simplified calculation that assumes an engine efficiency of 0.3 at all waypoints. (Increasing the assumed engine efficiency to 0.4 reduces $|A_f|/|A|$ from $0.58$ to $0.49$ but does not alter any high-level conclusions.)  We then use $A_f$ to define
\begin{equation}
    \epsilon_{form} = \frac{\sum_{A_f} EF_{cost} - EF_{contrail}}{\sum_{A_f} EF_{cost}} \,\, ,
\end{equation}
as the effectiveness of avoidance maneuvers that target formation and
\begin{equation}
    \epsilon_{other} = \frac{\sum_{A - A_f} EF_{cost} - EF_{contrail}}{\sum_{A - A_f} EF_{cost}} \,\, ,
\end{equation}
as the effectiveness of avoidance maneuvers that target other features. Finally, we note that $\epsilon_{form}$ and $\epsilon_{other}$ are related to the overall effectiveness by
\begin{equation}
    \epsilon = \epsilon_{form} \,f_{form} + \epsilon_{other} (1 - f_{form}) \,\, ,
\end{equation}
where
\begin{equation}
    f_{form} = \frac{\sum_{A_f} EF_{cost}}{\sum_A EF_{cost}}.
\end{equation}
is the fraction of cost-optimal EF produced by the subset of waypoints for which avoidance maneuvers target formation.

$\epsilon_{form}$ is close to 1 at all lead times (Figure \ref{fig:avoidance-success}a), indicating that avoidance maneuvers that target formation are highly effective when used. This is likely due to contrail formation depending primarily on temperature: temperature has relatively high predictability in the UTLS \cite{gierens_how_2020}, suggesting that flights re-routed into regions where temperatures are forecast to be too high for contrails to form are likely to end up in regions where temperatures are actually too high for contrails to form. In contrast, $\epsilon_{other}$ is slightly lower ($\sim 0.8$ to $1$) at short lead times and much lower ($\sim 0.5$) at 48 hour lead times, indicating that avoidance maneuvers that target features other than formation are significantly less effective. Overall effectiveness is high (i.e., closer to $\epsilon_{form}$ than $\epsilon_{other}$) because waypoints re-routed into regions that the forecast predicts will prevent contrail formation account for the majority ($\sim 60$\%) of the cost-optimal EF produced by all waypoints included in avoidance maneuvers. Together, these results suggest that overall effectiveness is high because avoidance based on formation is both frequent and effective. 

We emphasize that our optimizer is {\it not} deliberately attempting to avoid all regions where the SAC is met. Both cost- and contrail-optimal trajectories contain long segments through regions where flights form transient contrails, and the optimizer will not penalize these segments so long as they are not within an ISSR. It is only once a flight enters an ISSR and produces a persistent warming contrail that the flight segment will be penalized proportional to the predicted contrail radiative forcing. The contrail-optimal alternative is often to eliminate the persistent contrail by maneuvering (for a relatively short period of time) out of the region where the SAC is met. Avoiding {\it all} regions where the SAC is satisfied (regardless whether they support persistent contrails) would require aircraft to avoid all regions cold enough for transient contrails to form and would not be operationally feasible. 

\subsection{Implications for assessing forecast accuracy} \label{sec:speculation}

The preceding discussion is focused on how our optimizer accommodates forecast instability (i.e., errors in forecasts relative to reanalysis). However, the real-world effectiveness of navigational avoidance is also sensitive to forecast accuracy (i.e., to errors in forecasts or reanalysis relative to reality). Evaluating forecast accuracy is a much more challenging topic than evaluating forecast stability, and largely lies beyond the scope of this study.  However, the analysis presented above does provide some methods that may be useful for assessing forecast accuracy.

In Section \ref{sec:design:forecast_errors}, we showed that pointwise error rates in predictions of ISSR locations are similar in forecasts compared to reanalysis and reanalysis compared to reality (Figure \ref{fig:forecast_ets}). We also showed that proximity distributions, which characterize the spatial structure of forecast errors, are remarkably similar when comparing forecasts to reanalysis and reanalysis to IAGOS measurements. In particular, nearly all ($>90\%$) of waypoints inside IAGOS ISSRs are within an hour's flight time (either forward or backward along the current trajectory) of an ERA5 ISSR at the same altitude found in the aircraft trajectory, just as nearly all waypoints inside ERA5 ISSRs are within an hour's flight time of ISSRs in HRES forecasts. This indicates that errors in ISSR locations in forecasts relative to reanalysis and reanalysis relative to reality occur at similarly-small spatial scales, and suggests that an optimizer that accommodates small-scale errors in forecasts relative to reanalysis may also be capable of accommodating errors in forecasts relative to reality.

Our analysis of the effectiveness of avoidance maneuvers (Section \ref{sec:effectiveness}) requires information about pairs of cost- and contrail-optimal trajectories and so cannot easily be adapted to use measurements from IAGOS flights. However, one key result---that avoidance maneuvers that avoid formation are likely to be successful when considering errors between forecasts and reanalysis---may remain true when considering errors between forecasts and the real world. This is because forecasts of UTLS temperature are better-constrained by observations than forecasts of UTLS humidity and show relatively good agreement with IAGOS measurements \cite{gierens_how_2020, reutter2020ice}, and contrail formation is primarily temperature- rather than humidity-dependent. Further, recent experiments and modeling suggests that higher ambient temperatures significantly reduce contrail impacts from lean-burn engines \cite{ponsonby2025updated}. This suggests that updated contrail models that more accurately account for vPM activation may in turn exhibit increased forecast skill since temperature is both a stable and accurate forecast feature.
It is worth noting that these findings do not imply that performing navigational avoidance based on formation criteria alone would be advisable, as such a strategy would result in prohibitively large fuel penalties.

\subsection{Operational Implications}

We note that in practice, navigational avoidance of persistent contrails will involve additional fuel burn and changes in flight times.  In \ref{sec:fuel}, we present a detailed discussion of the penalties associated with the avoidance profiles used in this study. Specifically, avoidance of approximately 80\%-90\% of contrail energy forcing is achieved with an approximate fuel penalty between 0.2\% and 0.3\% additional fuel burn depending on the choice of minimum segment length.  While this number is not insignificant, we note that more intelligent optimization procedures (e.g., strategies using lateral as well as vertical avoidance) may substantially reduce this penalty (see \ref{sec:fuel} for futher discussion). Other operational considerations, such as effects on workload, airspace congestion, interactions with other meteorological safety considerations (e.g., turbulence and icing), are important but left to future work. 

Finally, we note that we have performed our simulations with a single cost index setting. In practice, operators plan trajectories at a variety of cost indices based on various economic and operational considerations. An important future piece of work would be to perform a range of contrail avoidance simulations at varying cost indices to understand the effects of variable cost indices on the cost penalties and efficacy of navigational contrail avoidance.  Such a study may be best carried out by a commercial grade flight planning system similar to \cite{frias_feasibility_2024}.

\section{Conclusion} \label{sec:conclusion}

Our results suggest that forecasts of contrail EF at lead times relevant to flight planning ($< 48 \,\mathrm{hrs}$) are sufficiently stable to support navigational contrail avoidance in the flight planning stage. Contrail-optimal trajectories generated based on HRES forecasts with lead times of up to 6 hours reduce contrail radiative forcing assessed using ERA5 reanalysis by amounts exceeding 90\%. Trajectory optimization using forecasts with longer lead times most relevant for flight planning (8-24 hrs) is slightly less effective, but can still reduce reanalysis-assessed EF by 80-90\%. Our methods---trajectory optimization that attempts to reduce contrail warming using forecasts generated with a simplified contrail process model, followed by a ``retrospective'' assessment using reanalysis meteorology and a high-fidelity contrail process model---intentionally mimic a workflow that could be used by operators to manage and report contrail impacts. Our results demonstrate that, within such a workflow, the reported reduction in contrail warming is unlikely to be strongly sensitive to the lead time of the forecast used for flight planning.

Our results further show that forecast-based trajectory optimization can eliminate over 90\% of the total contrail EF assessed using ERA5 in spite of frequent pointwise differences in locations of ISSRs and high-EF regions ($\mathrm{ETS} < 0.4$). We suggest that the robustness to pointwise forecast errors is linked to the way that the minimum segment length of our optimizer interacts with the spatial scale of the forecast errors. A sufficiently long minimum segment length prevents the optimizer from fitting trajectories closely to small-scale variations in forecast EF, which often disagree with small-scale variations in the reanalysis EF, and instead requires the optimizer to avoid broader envelopes that contain pockets of high EF in both forecasts and reanalysis. In essence, the relatively simple constraint imposed via the minimum segment length encourages the optimizer to compute contrail-optimal trajectories that take advantage of useful information in forecasts without overfitting to noise.

Our results highlight the potential value of designing robust optimizations strategies that accommodate uncertainties in contrail forecasts. This study's approach to optimizer design was labor-intensive: it was based on intuition about the spatial structure of forecast errors developed by looking at maps of contrail forecasts and reading papers about the morphology of contrail-forming regions \cite{wolf_2024_distribution}, and translating this intuition into an optimization strategy required hand-tuning an optimizer parameter based on a manual analysis of forecast errors. Moving toward a more formal representation of forecast uncertainty (e.g., based on probabilistic or ensemble forecasts) could eliminate the need to hand-tune optimizers. However, more research is needed to determine how best to represent uncertainty in contrail forecasts designed for use in flight planning systems.

There are a number of important questions not addressed by this study. First and foremost, our work does not directly address the critical question of the accuracy of modeled contrail impacts relative to reality. Our analysis of the way our optimizer interacts with errors in forecasts relative to reanalysis, together with analysis showing key similarities in the frequency of errors in reanalysis relative to reality, provides some reason for optimism that current forecast models may provide a better foundation for navigational avoidance than previous studies \cite{gierens_effect_2022,hofer_how_2024} have suggested. While the sparsity of UTLS temperature and humidity observations prevents the holistic approach used to assess forecast stability from being applied to forecast accuracy, this study strongly suggests that pointwise error metrics provide (at best) an incomplete or (at worst) a misleading measure of the suitability of forecasts for navigational avoidance. That said, our study in no way provides affirmative evidence that contrail forecasts are accurate enough for navigational avoidance to provide a climate benefit. Determining whether they are requires using observations to assess the impact of re-routing flights, and that is a step that this paper does not take.

The accuracy of modeled contrail impacts is also dependent on the fidelity of process models used to estimate radiative forcing. CoCiP has been tuned for agreement with some key observations and high-fidelity models \cite{schumann_contrail_2012, unterstrasser_properties_2016}, but nevertheless includes sources of both structural and parametric uncertainty. Parametric uncertainty can be quantified, to an extent, with perturbed parameter ensembles \cite{engberg_2025_forecasting,platt_effect_2024}. However, structural uncertainty is currently difficult to quantify given the lack of diversity in computationally-inexpensive contrail process models.

Finally, our work assumes that flights follow optimized trajectories exactly. This will not be true in practice: delays or other operational issues (e.g., the need to avoid turbulence, or traffic-related restrictions) may cause aircraft to deviate from contrail-optimal flight plans. Additional factors, such workload considerations by pilots or dispatchers, or other human factors, may further limit the efficacy of navigational avoidance. In practice, implementation of navigational avoidance will necessitate some amount of tactical decision making by pilots, dispatchers, and air-traffic managers. The impact of these deviations on contrail EF has not been quantified here but will be a focus of future studies and operational trials.

\section*{Acknowledgments}
IAGOS data were created with support from the European Commission, national agencies in Germany (BMBF), France (MESR), and the UK (NERC), and the IAGOS member institutions (http://www.iagos.org/partners). The participating airlines (Lufthansa, Air France, Austrian, China Airlines, Hawaiian Airlines, Air Canada, Iberia, Eurowings Discover, Cathay Pacific, Air Namibia, Sabena) supported IAGOS by carrying the measurement equipment free of charge since 1994. The data are available at http://www.iagos.fr thanks to additional support from AERIS. This publication contains modified information from the Copernicus Climate Change Service and data and products of the European Centre for Medium-Range Weather Forecasts (ECMWF). Neither the European Commission nor ECMWF is responsible for any use that may be made of the Copernicus information or data it contains. This work was presented, in part, at the 2024 WMO AeroMetSci Meeting in Geneva, Switzerland. TD, TA, ZE, NM, and MLS are employees of Breakthrough Energy, a collection of for-profit and not-for-profit activities committed to bringing the world to net-zero emissions by 2050. RT, JPI, and MEJS are funded, in part, by Breakthrough Energy.

\appendix

\section{Fuel and Delay Penalties} \label{sec:fuel}

In this appendix, we describe the penalties in flight time and fuel burn incurred by contrail-optimal trajectories generated by our optimizer.
To highlight the flexibility of the optimization procedure used in this work, we show that the total fuel penalty can be reduced by adjusting the contrail cost index. 
We note that the focus of this study is to evaluate the impact of forecast instability on the effectiveness of contrail avoidance.  
As such, the numbers reported here should not be interpreted as the true cost of contrail avoidance. More sophisticated optimization procedures may be able to significantly reduce cost increases.

 \begin{figure}[t!]
    \centering
    \begin{subfigure}{0.485\linewidth}
        \includegraphics[width=\linewidth]{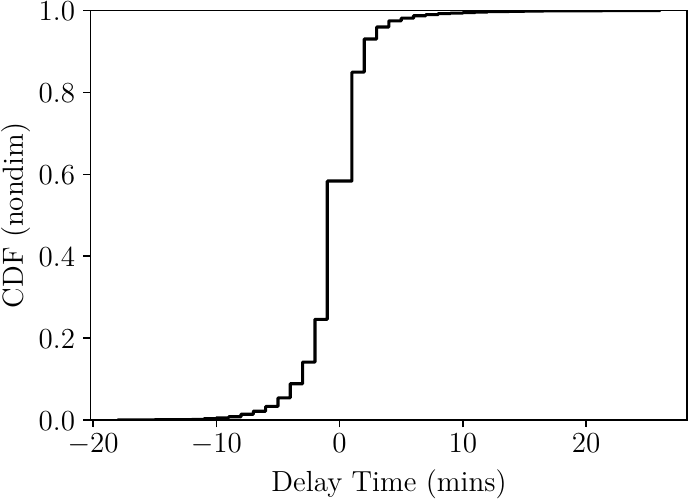}
        \caption{}
        \label{fig:delay}
    \end{subfigure}
    \hfill
    \begin{subfigure}{0.485\linewidth}
        \includegraphics[width=\linewidth]{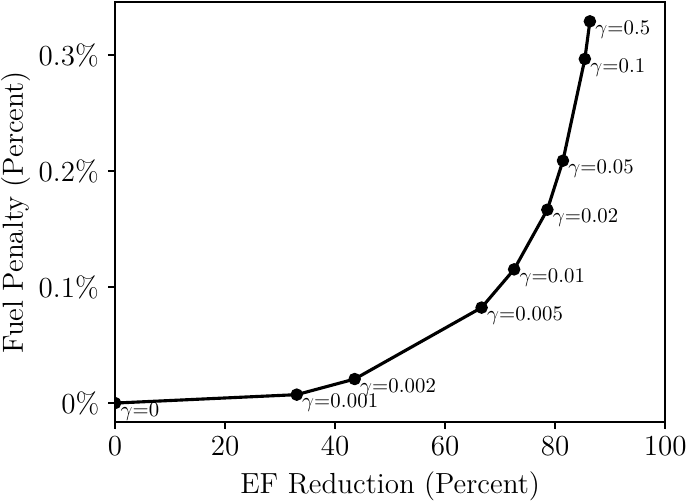}
        \caption{}
        \label{fig:fuel}
    \end{subfigure}
    \hfill
    \begin{subfigure}{0.45\linewidth}
        \includegraphics[width=\linewidth]{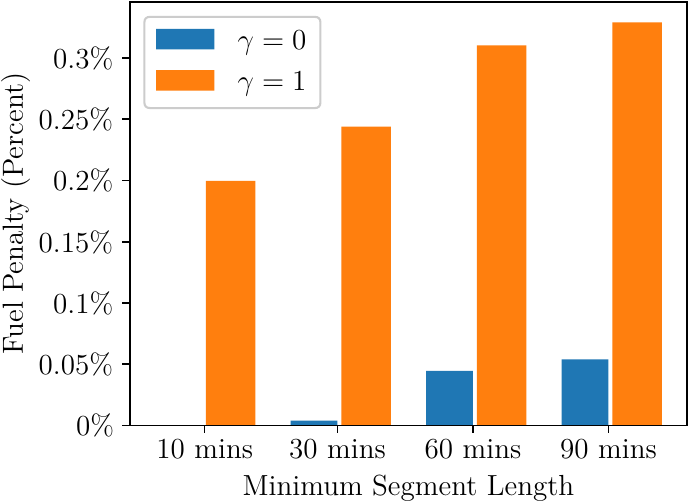}
        \caption{}
        \label{fig:fuel_seg_len}
    \end{subfigure}
    \caption{(a) The empirical cumulative distribution of changes in flight times for contrail avoidance flights. The majority ($>58$\%) of avoidance flights are of equal or shorter in duration due to the fact that they almost always fly at lower altitudes with higher true air speeds. (b) Fleet-wide fuel penalties for different contrail cost indexes and the associated contrail energy forcing achieved with each. Here, $\gamma$ represents a scaling factor applied to the value of 6.7e-11 kg/J used for the results in the main body of the paper. 
    (c) Fleet-wise fuel penalties for different minimum segment lengths considering cost optimal ($\gamma=0$), and contrail optimal ($\gamma=1$). Imposing a 90-minute minimum segment length constraint produces a fuel burn penalty between 0.05\% and 0.1\% compared to a baseline obtained using the $\gamma=0$ and a 10-minute minimum segment length.
    }
\end{figure}

The distribution of changes in flight times between contrail and cost optimal routes is depicted in Figure \ref{fig:delay}. 
Most contrail avoidance flights are shorter in duration because avoidance is usually performed by descending to a lower flight level. 
For a fixed Mach number, the true airspeed is higher at a lower altitude, leading to a shorter flight duration. 
Across all fights considered in this study, including those not re-routed due to an absence of contrails, contrail-optimal flight times were on average 5 seconds shorter. Flights that were re-routed were on average 26 seconds shorter.
Of the avoidance flights, 58\% were of shorter duration than the cost-optimal flight plan, and 85\% were within one minute of the original flight duration. 
In practice, a shorter average flight duration would offset some of the operational costs of navigational avoidance for the operator.

Example fuel burn penalties are depicted in Figure \ref{fig:fuel}.  This figure depicts the trade-off between excess fuel burn and contrail EF reduction, obtained by adjusting the contrail cost index. 
Here, $\gamma$ is used as a scaling factor applied the value of 6.7e-11 kg/J used in the results presented in the main body of this paper; that is, $\gamma = 1$ indicates a contrail cost index of 6.7e-11, and $\gamma = 0$ indicates the cost-optimal route.  
The values in Figure \ref{fig:fuel} were obtained by optimizing flights against only the forecast from 12Z the day before flight departure (lead times of 12-36 hours) rather than the entire set of forecasts used previously.  
We note that fuel penalties associated with avoidance strategies are unlikely to be sensitive to forecast lead times.

By examining Figure \ref{fig:fuel}, we see that vertical deviations can produce a 86\% EF reduction with 0.33\% additional fuel burn.  
The near vertical slope of the line at the right-hand side of the figure suggests that further increases to the contrail cost index would not achieve significant additional EF reduction.  
On the left-side side of the figure, it can be seen that over 33\% EF reduction can be achieved with less than 0.01\% additional fuel burn if the cost index is decreased by a factor of 1000. 

In Figure \ref{fig:fuel_seg_len} we examine the fuel burn penalty caused by the minimum segment length constraint of the optimizer.  
We note that for nearly all trajectories in this study, the spacing used in the trajectory optimization search procedure has a segment length greater than 10 minutes, and so the 10-minute minimum segment length constraint may be interpreted as no constraint.
In this figure, we see that at a contrail cost index of 0, the 90-minute constraint produces approximately a 0.05\% fuel burn penalty, and at a contrail cost index of 1, we see approximately a 0.1\% fuel burn penalty compared to the case of the 10-minute minimum segment length. We note that the 10-minute minimum segment length constraint may be unrealistically short for use in commercial flight planning, and that commercial-grade optimizers typically constrain the number of step-climbs by imposing engine-cycle costs which penalize changes in thrust, rather than constraining the time between step climbs.

We again re-iterate that we provide the fuel burn estimates here as very rough approximations to the cost of contrail avoidance, and highlight that the optimization procedure presented in this work is highly flexible. As stated previously, we believe a more sophisticated optimizer (particularly one that considers lateral as well as vertical deviations) may be able to achieve similar EF reductions with a smaller fuel penalty. This is a topic of forthcoming research.

\section{Monthly Variability}
\label{sec:monthly}

 \begin{figure}[t!]
    \centering
    \begin{subfigure}{0.485\linewidth}
        \includegraphics[width=\linewidth]{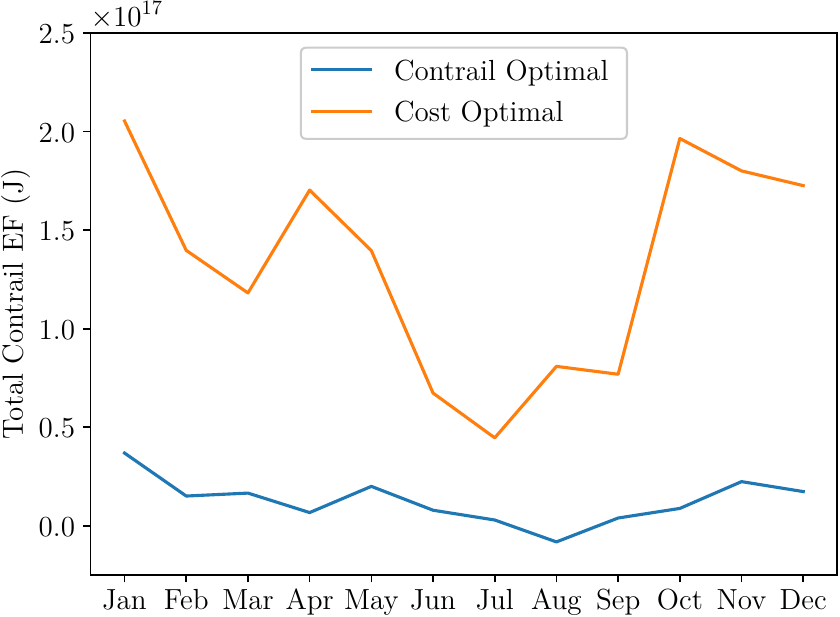}
        \caption{}
        \label{fig:ef_by_month}
    \end{subfigure}
    \hfill
    \begin{subfigure}{0.485\linewidth}
        \includegraphics[width=\linewidth]{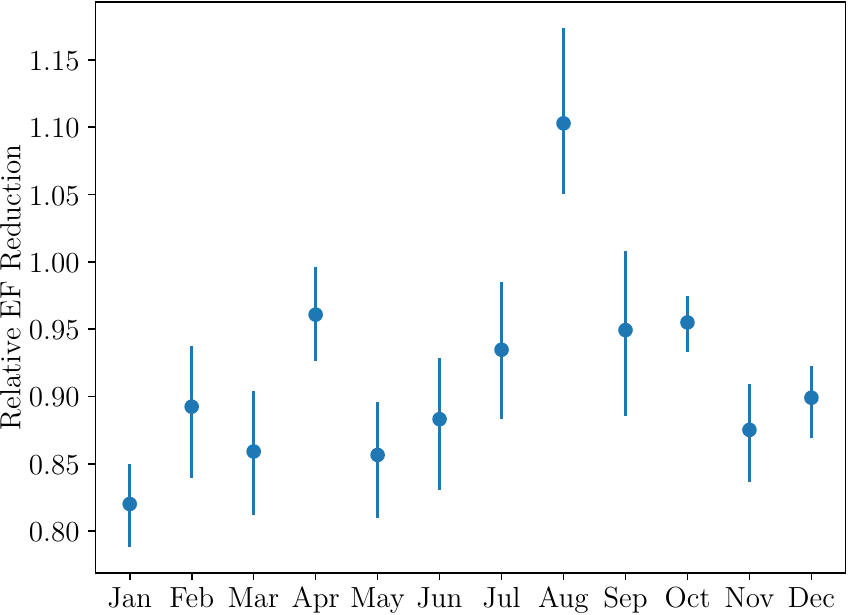}
        \caption{}
        \label{fig:cis_by_month}
    \end{subfigure}
    \caption{(a) The total fleet-wide contrail energy forcing for each day sampled in the study. One day per month was chosen at random and 5000 flights were simulated on each day.  The impact of the cost optimal trajectories is shown in orange which exhibits expected seasonal variability. The resulting contrail impact after optimization according to a short-range forecast is shown in blue. (b) The percentage of contrail warming removed by month, with 95\% confidence intervals. Apart from one outlier, the results are well clustered around the mean, and all months obtain greater than 80\% reduction.
    }
\end{figure}

\begin{figure}
    \centering
    \includegraphics[width=0.6\linewidth]{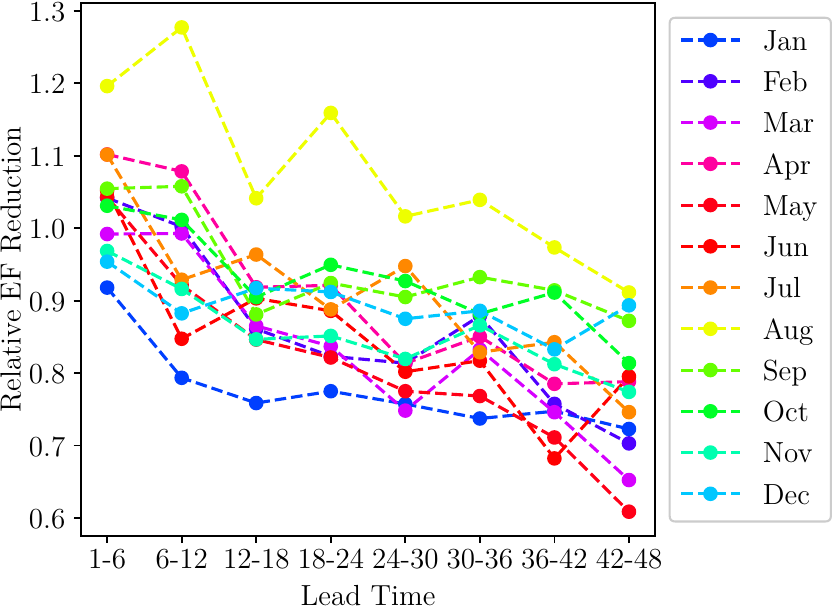}
    \caption{Change in fleet-wide contrail EF as a function of forecast lead time by month. While there is some spread in total reduction by month, all months follow the same overall trend of forecast stability, with a reduction of mitigated EF of 10\%-20\% as lead times increase from 1 to 48 hours}
    \label{fig:monthly_means}
\end{figure}

In this appendix, we consider monthly trends of contrail warming reduction.  In this study, we sampled one day per month, and for each day simulated a total of 5000 flights. We chose to sample one day per month at random rather that a random sampling of 12 days in order to sample a wide range of meteorological regimes. In Figure \ref{fig:ef_by_month} we show the total fleet-wide contrail impact for each day sampled in the study. Here, we depict both the impact of the cost optimal trajectories, along with the residual contrail impacts for flights optimized to mitigate contrail impacts. The flights depicted in this figure were optimized with a minimum segment length of 90 minutes using a short-range forecast with a maximum lead time of 24 hours. Contrail impacts were evaluated using ERA5 reanalysis and the trajectory version of CoCiP. Figure \ref{fig:cis_by_month} shows the same data, but depicts the relative EF reduction between the two sets of flight trajectories.  The error bars depict the 95\% confidence interval obtained by bootstrapping per-flight. The results are well clustered around the sample mean of 90.5\% [89.4\%, 91.7\%], with no month achieving an EF reduction below 80\%. We note that in outlier month of August, the large fractional reduction in contrail EF does not much change the 12 month average results because of the relatively small amount of EF avoided in August.

In Figure \ref{fig:monthly_means}, we show the EF reduction achieved when binning the results by forecast lead time. This figure shows the same data depicted in Figure \ref{fig:ef_reduction_grid_traj} for each individual month in the study.  While there is some variability in the total mitigation rate by month, as previously depicted in Figure \ref{fig:cis_by_month}, the overall trend by lead time for each month shows very little variability. For all months, the total reduction in EF at 36 hour lead times is less than 20\% lower than the reduction at short lead times. We note that there is slightly more variability in the amount of mitigation obtained with lead times longer than 36 hours, with some, but not all, months exhibiting further degradation.

These analyses suggest that the central result claimed in this work---that, compared to reanalysis, 80\%-90\% of contrail warming may be mitigated through the flight planning using forecasts with lead times of 8-24 hours---is unlikely to change by sampling additional days.

\section*{References}

\bibliographystyle{unsrt}
\bibliography{references}

\end{document}